\UseRawInputEncoding
\pdfoutput=1

\documentclass{article}
\usepackage{spconf,amsmath,graphicx}
\usepackage{cite}
\usepackage{amssymb,amsfonts}
\usepackage{framed,multirow}
\usepackage{subfigure} 
\usepackage{algorithm}
\usepackage{algorithmic}
\usepackage{textcomp}

\title{Spatio-Temporal Structure Consistency for Semi-supervised Medical Image Classification}                                                                                                                                                                                    
\name{Wentao Lei\textsuperscript{1,$\dagger$}, Lei Liu\textsuperscript{1,$\dagger$}, Li Liu\textsuperscript{2,*} 
\thanks{* Corresponding author: avrillliu@hkust-gz.edu.cn.}\thanks{$\dagger$ Equal contribution.}}
\address{\textsuperscript{1}Shenzhen Research Institute of Big Data, The Chinese University of Hong Kong, Shenzhen, China\\
\textsuperscript{2}The Hong Kong University of Science and Technology (Guangzhou)}

\begin{document}
\maketitle
\begin{abstract}
Intelligent medical diagnosis has shown remarkable progress on the large-scale datasets with full annotations. 
However, very few labeled images are available due to significantly expensive annotations by experts.
To efficiently leverage abundant unlabeled data, we propose a novel Spatio-Temporal Structure Consistent (STSC) learning framework to combine both spatial and temporal structure consistency. Specifically, a gram matrix is derived to capture the structural similarity among different training samples in the representation space. At the spatial level, our framework explicitly enforces the consistency of structural similarity among different samples under perturbations. At the temporal level, we desire to maintain the consistency of the structural similarity in different training iterations by digging out the stable sub-structures in a relation graph. Experiments on two medical image datasets (\textit{i.e.}, ISIC 2018 and ChestX-ray14) show that our method outperforms state-of-the-art Semi-Supervised Learning (SSL) methods. Furthermore, extensive qualitative analysis on the Gram matrices and heatmaps by Grad-CAM are presented to validate the effectiveness of our method.

\end{abstract}
\begin{keywords}
Medical Image Classification, Semi-supervised Learning, Spatial-temporal Structure Consistency
\end{keywords}
\section{Introduction}
\label{sec:intro}
Deep learning has achieved success in medical image analysis with large-scale datasets and manual annotations. Studies\cite{DBLP:journals/corr/abs-1712-00409,liu2020re} have shown that larger labeled datasets lead to better performance. However, obtaining accurate labels for medical images is costly and time-consuming due to the need for clinical expertise, making it challenging to build a large-scale dataset with accurate labels\cite{liu2020semi,chen2021uscl,chen2022generating}.

To alleviate the over-dependence on annotations, SSL method is designed to improve performance by leveraging abundant unlabeled data with limited labeled data.
SSL approaches can be roughly categorized into two categories: self-training \cite{doi:10.1080/01621459.1975.10479874} and consistency regularization \cite{DBLP:journals/corr/LaineA16}. The former generates artificial pseudo labels from the predictions of unlabeled data and then adds them into the training set \cite{Xie_2020_CVPR}. The latter improves the availability of unlabeled data based on the predictions consistency between different modified versions of the same input \cite{NIPS2016_30ef30b6,NIPS2014_66be31e4,wang2021self}. As one of the key methods of SSL, consistency regularization aims to capture the relationships for both labeled and unlabeled data in the feature space. For instance, \cite{NIPS2017_68053af2} enforced the prediction consistency of the same sample under different perturbations to learn a unified feature space. \cite{9095275} emphasized the consistency of the spatial relationships among different samples via the relation information of unlabeled data. Above-mentioned approaches mainly investigate the spatial structure relationships to utilize unlabeled samples. However, there is a lacking of effective exploration for temporal consistency, which could maintain stable spatial relationships along with training.

\begin{figure*}[!t]
\centerline{\includegraphics[width=0.75\textwidth]{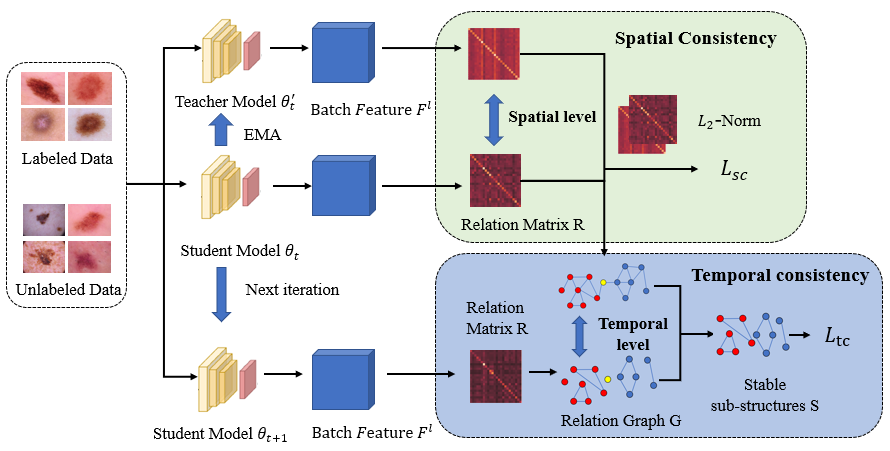}}
\caption{The pipeline of the proposed STSC. The teacher model is updated as the EMA of the student model. Spatio-temporal consistency consists of two parts: spatial relation loss $L_{sc}$ and temporal relation loss $L_{tc}$.}
\label{fig1}
\end{figure*}

Medical professionals frequently consult prior samples to aid in their diagnostic decision-making. The value of these samples increases with higher confidence. Drawing inspiration from this clinical practice, we propose a novel spatio-temporal structure consistent (STSC) semi-supervised framework (see Fig. \ref{fig1}) to explore spatial and temporal structural relationships among different samples simultaneously. In particular, a case-level gram matrix is derived to describe the similarity among the different samples in the representation space. Then we transform the gram matrix into an adjacency matrix to represent the graph structure of training samples. During training, stable spatial structure can be obtained by encouraging consistency of gram matrix under different perturbations on the inputs. At the late training stage, we present a Temporal Sub-structure Consistency (TSC) method to maintain the temporal consistency of the structural relationships, which further captures the stable sub-structures in a relation graph along with training. More discriminative semantic information can be learned from the relationships of unlabeled data guided by spatio-temporal structure consistency.


In summary, the main contributions are: \textbf{(1)} A novel STSC semi-supervised learning framework is proposed to efficiently leverage the unlabeled data, which reduces the requirement of labeled data on both single-label and multi-label tasks. \textbf{(2)} We propose a Temporal Sub-structure Consistency (TSC) method to explore the stable sub-structures in a relation graph. It can effectively capture the stable sample structure along with training. \textbf{(3)} Experiments on two public medical image datasets (\textit{i.e.}, ISIC 2018 and ChestX-ray14) demonstrate the superior performance of our approach compared with the state-of-the-art (SOTA) methods.

\section{Method}
\label{section:method}


\subsection{Spatio-Temporal Structure Consistent Framework}
\textbf{Preliminaries.} We consider the basic image classification task with $c$ categories. Under semi-supervised setting, we have a labeled set $\mathcal{D}_L$ = $\{(x_i, y_i)\}_{i=1}^{N}$ with $N$ labeled samples, where $y_i$ represents the label for $x_i$. There is also an unlabeled set $\mathcal{D}_U$ = $\{x_i\}_{i=N+1}^{N+M}$ with $M$ unlabeled samples. The target is to learn a deep model $f(\cdot, \theta)$ parameterized by $\theta$ based on both labeled and unlabeled training set. Typically, the consistency regularization based semi-supervised learning aims to optimize the following objective function:
\begin{equation}
\min _{\theta} \!\! \sum_{(x,y) \in \mathcal{D}_L} \!\!\!\!\!\mathcal{L}_{s}\left(f\left(x, \theta\right), y\right)+\lambda \!\!\!\!\!\!\! \sum_{x \in \mathcal{D}_U \cup \mathcal{D}_L} \!\!\!\!\!\! \mathcal{L}_{u}\left(f(\eta, x, \theta), f(\hat{\eta},x,\theta^{\prime})\right),
\label{eq1}
\end{equation}
where $\mathcal{L}_{s}$ denotes the supervised loss on the labeled set $\mathcal{D}_L$, and $\mathcal{L}_{u}$ is the unsupervised consistency loss to force the consistent predictions of the same inputs under different perturbations. Here, we exploit the teacher-student structure with the same network architecture, parameterized by $\theta$ and $\theta^{\prime}$, respectively. $\eta$ and $\eta^{\prime}$ denote the different perturbations applied to the same input images. $\lambda$ is a hyper-parameter that controls the trade-off between the supervised and unsupervised loss.

\noindent\textbf{Teacher-Student Model.}  The teacher model is updated as the exponential moving average (EMA) of the weights $\theta$ of the student model. The teacher parameters $\theta_t^{\prime}$ are updated as :
$\theta_{t}^{\prime}=\alpha \theta_{t-1}^{\prime}+(1-\alpha) \theta_{t}$
at training iteration $t$. $\alpha$ is a hyper-parameter to control the updating rate. Our framework preserves the conventional individual consistency mechanism to force the consistent predictions of teacher and student model \cite{DBLP:journals/corr/LaineA16} under different perturbations, which optimizes the following sample-level loss:
\begin{equation}
\mathcal{L}_{c}=\sum_{x_i \in \mathcal{D}_U \cup \mathcal{D}_L}
\mathbb{E}_{\eta^{\prime},\eta}
\left \| f\left(x_{i},\theta^{\prime}, \eta^{\prime}\right)-f
\left(x_{i}, \theta, \eta\right)\right\|_{2}^{2}.
\end{equation}
\subsection{Spatial and Temporal Structure Consistency}
In this section, we first derive a case-level Gram Matrix \cite{DBLP:journals/corr/GatysEB15a} to capture the structural relations among different samples. Given a mini-batch with $B$ samples, let $F^l \in R^{B\times H\times W\times C}$ denote the metrics of activation maps at layer $l$, where $H$ and $W$ are the sizes of the feature map, and $C$ is the channel number. The feature map $F^l$ is first reshaped as $D^l \in R^{B\times HWC}$. Then the Case-wise Gram Matrix $M^l \in R^{B\times B}$ is computed as: $M^{l}=D^{l} \cdot\left(D^{l}\right)^{T}$,
where $M_{ij}$ is the inner product between $D^{l}_{(i)}$ and $D^{l}_{(j)}$ to measure the similarity between the activations of $i^{th}$ and $j^{th}$ sample in the mini-batch. The final sample relation matrix
$R_l$ is obtained by conducting the $L_2$ normalization for each
row $M^l_i$ of $M^l$, which is expressed as:
\begin{equation}
R^{l}=\left[\frac{M_{1}^{l}}{\left\|M_{1}^{l}\right\|_{2}}, \ldots, \frac{M_{B}^{l}}{\left\|M_{B}^{l}\right\|_{2}}\right]^{T}.
\end{equation}
The spatial structure consistency regularizes the relation matrix $R^l$ to be stable under different perturbations, preserving the spatial semantic relation
among samples. The proposed spatial structure consistency loss is defined as:
\begin{equation}
\mathcal{L}_{sc}=\!\!\sum_{\mathcal{X} \in\left\{\mathcal{D}_{U} \cup \mathcal{D}_{L}\right\}} \!\!\frac{1}{B}\left\|R^{l}(\mathcal{X} ; \theta, \eta)-R^{l}\left(\mathcal{X} ; \theta^{\prime}, \eta^{\prime}\right)\right\|_{2}^{2}.
\end{equation}

\subsection{Temporal Sub-structure Consistency}
To further explore the structure of the training samples at the temporal level, we propose a Temporal Sub-structure Consistency to maintain stable spatial information based on a graph. 

Concretely, we first obtain an adjacent matrix $A$ by binarizing the Gram Matrix $R$. Then a threshold $\tau$ is used to filter sample edges with weak semantic relationship. The element $A_{(i,j)}$ will be set as one if its corresponding position in Gram Matrix is larger than $\tau$, indicating $x_i$ and $x_j$ have closer semantic relationship, otherwise $A_{(i,j)}$ will be set to zero:
\begin{equation}
\label{eq6}
A_{(i,j)}=\left\{
\begin{aligned}
1,  \quad R_{(i,j)}\geq \tau, \\
0,  \quad R_{(i,j)} < \tau.
\end{aligned}
\right.
\end{equation}


A graph $G(X, E)$ is generated from the adjacent matrix $A$, where $X$ is the set of vertices (samples) and $E$ is the set of edges (edge $E(x_i, x_j)$ exists when $A(i,j) = 1$). During training, we identify stable sub-structures $s(\bar{X})$ in $G$ where all elements in $\bar{X}$ are connected in both time $t$ and $t+1$. The collection of stable sub-structures is denoted as $\mathcal{S} = \{s_i\}_{i=1}^{k}$, where $s_i$ is the $i^{th}$ stable sub-structures and $k$ is the amount of stable sub-structures.

The temporal structure consistency is designed to maintain stable sub-gram matrix $R^l$ at different training iterations, which obtains temporal relationships among of different samples. The temporal structure consistency is defined as:
\begin{equation}
\mathcal{L}_{tc}=\!\!\sum_{i = 1}^k\sum_{\mathcal{X} \in\left\{s_{i}\right\}} \!\!\frac{1}{B}\left\|R^{l}_{t}(\mathcal{X} ; \theta, \eta)-R^{l}_{t-1}\left(\mathcal{X}; \theta, \eta \right)\right\|_{2}^{2}.
\end{equation}

Finally we can obtain the total objective functions for the STSC semi-supervised framework as follows:
\begin{equation}
    \begin{aligned}
        & \mathcal{L}=\mathcal{L}_{s}+\lambda \mathcal{L}_{u}, 
        & \mathcal{L}_{u}=\mathcal{L}_{c}+\beta \mathcal{L}_{sc}+\gamma \mathcal{L}_{tc},
    \end{aligned}
\end{equation}
where the supervised objective and the unsupervised objective are denoted as $L_s$ and $L_u$ respectively. The unsupervised objective is composed of the individual consistency $L_c$ and the spatial-temporal consistency $L_{sc}$ and $L_{tc}$. To keep a balance between $L_c$, $L_{sc}$ and $L_{tc}$, we set 2 hyperparameters denoted as $\beta$, $\gamma$, which are generally set as 1, and $\lambda$ is the trade-off weight between the supervised and unsupervised loss.


\section{Experiment}
\subsection{Experimental Setup}
\textbf{Datasets.} The experiments are conducted on two public medical datasets: ChestX-ray14\cite{DBLP:journals/corr/WangPLLBS17} and ISIC 2018 Skin Lesion Analysis \cite{DBLP:journals/corr/abs-1803-10417,DBLP:journals/corr/abs-1902-03368}. For fair comparisons, ISIC 2018 data is randomly split into a training, a validation and a test set ($7:1:2$) following \cite{9095275}. ChestX-ray14 dataset is split into training, validation, and test sets ($7:1:2$) following \cite{avilesrivero:hal-02193970}.

\noindent \textbf{Implementations Details.} For ISIC2018 dataset, the batch size is $64$. For ChestX-ray14 dataset, the batch size is $48$. The Adam is used with initial learning rate ($1e-4$), which is decayed by a power of $0.9$ after every epoch. All experiments are conducted on 4 tesla-V100 GPUs. The evaluation metrics include AUC, Accuracy, Sensitivity and Specificity.

\subsection{Result and Analysis}
\textbf{ISIC 2018 Dataset.} In Tab. \ref{table1}, we compare the performance of the proposed method with previous approaches using 20\% labeled data on the ISIC 2018 dataset. The upper bound performance is obtained as a baseline by training a supervised model using 100\% labeled data. The self-training method obtains higher specificity than the other approaches, which benefits from negative samples. Compared with previous SOTA method SRC-MT, our method can outperform about 3.58\%, 1.78\% and 7.10\% for the AUC, accuracy, and sensitivity, respectively. Notably, our approach achieves improvements on all metrics over SRC-MT by enforcing the consistency of the spatial and temporal relationships among different samples, indicating the effectiveness of our method.

\begin{table}[!t]
\small
\centering
\caption{Comparison results on ISIC2018 dataset.}
\label{table1}
\renewcommand\tabcolsep{4.0pt} 
\begin{tabular}{c|c|c|ccccc}
\hline
\multirow{2}{*}{Method} & \multicolumn{2}{c|}{Ratio (\%)}   & \multicolumn{5}{c}{Metrics (\%)} \\\cline{2-8}
            &$\mathcal{D}_L$  &$\mathcal{D}_U$    & Acc  & Sen  & Spec  & AUC & F1\\\hline
\multirow{1}{*}{Upper}   & \multirow{1}{*}{100}& 0   & 95.10 & 75.20 & 94.94   & 95.43 &70.13 \\
\multirow{1}{*}{Baseline}      & \multirow{1}{*}{20} & 0   & 92.17 & 65.50 & 91.83   & 90.15 &52.03\\ \hline
\multirow{1}{*}{ST}\cite{Xie_2020_CVPR} & \multirow{1}{*}{20} & 80  & 92.37 & 67.63 & \textbf{93.31} & 90.58 &54.51  \\
\multirow{1}{*}{DCGAN}\cite{dcgan}      & \multirow{1}{*}{20} & 80  & 92.27 & 67.72 & 92.56   & 91.28 &  54.10\\ 
\multirow{1}{*}{TCSE}\cite{tcse}          & \multirow{1}{*}{20} & 80  & 92.35 & 68.17 & 92.51  & 92.24 &  58.44\\
\multirow{1}{*}{TE}\cite{DBLP:journals/corr/LaineA16}            & \multirow{1}{*}{20} & 80  & 92.26 & 69.81 & 92.55 & 92.70 & 59.33 \\ 
\multirow{1}{*}{MT}\cite{NIPS2017_68053af2}            & \multirow{1}{*}{20} & 80  & 92.48 & 69.75 & 92.20   & 92.96 & 59.10\\\hline
\multirow{1}{*}{SRC-MT}\cite{9095275}        & \multirow{1}{*}{20} & 80  & 92.54 & 71.47 & 92.72  & 93.58 & 60.68\\
\multirow{1}{*}{STSC (ours)}    & \multirow{1}{*}{20} & 80  & \textbf{93.95} & \textbf{72.60} & 92.94 & \textbf{93.73} &\textbf{60.84}\\\hline
\end{tabular}
\vspace{-0.3cm}
\end{table}

\begin{table}[!t]
\begin{center}
\caption{Comparison results on ChestX-Ray14 dataset.}
\label{table2}
\begin{tabular}{c|c|c|c|c|c}
\hline
Ratio & 2\% & 5\% & 10\% &15\% &20\% \\
\hline
GraphX \cite{avilesrivero:hal-02193970} & 53 &58 &63 &68 &78 \\
\hline
SRC-MT & \textbf{66.95} &\textbf{72.29} &75.28 &77.76 &79.23 \\
\hline
STSC (Ours) & 65.37  & 71.49 & \textbf{76.21} & \textbf{78.31}  & \textbf{79.45} \\
\hline
\end{tabular}
\end{center}
\vspace{-0.8cm}
\end{table}

\noindent\textbf{ChestX-Ray14 Dataset.} In Tab. \ref{table2}, we report the performance of our method and previous approaches under different percentages of labeled data on the ChestX-Ray14 dataset. $GraphX^{NET}$ is the baseline model and can achieve 78\% AUC using 20\% labeled data. However, its performance exhibits a large variance with respect to different labeled data percentages. Our method presents a more stable performance along with the changed labeled data percentages. SRC-MT is the previous SOTA method. Furthermore, it is indicated that the AUC of the STSC method is lower than SRC-MT method when the labeled data settings are 2\% and 5\%, showing superior performance when the labeled data percentage is increased. This phenomenon may be attributed to that the labeled data benefits from a more reliable relation structure.

\begin{table}[!t]
\small
\begin{center}
\caption{Performance on ISIC 2018 dataset.}
\label{table3}
\renewcommand\tabcolsep{4.0pt} 
\begin{tabular}{c|c|c|ccccc}
\hline
\multirow{2}{*}{Method} & \multicolumn{2}{c|}{Ratio (\%)}   & \multicolumn{5}{c}{Metrics (\%)} \\\cline{2-3}\cline{4-8}
&$\mathcal{D}_L$  &$\mathcal{D}_U$     & Acc. & Sen.  & Spec.   & AUC&F1\\ \hline
\multirow{1}{*}{Baseline} & \multirow{1}{*}{10} & 0    & 87.45 &64.22 &89.88 & 87.04 & 44.43\\            
\multirow{1}{*}{SRC-MT} & \multirow{1}{*}{10} & 90     &89.30   & 66.29 &90.47   &90.31 &  50.02\\
\multirow{1}{*}{STSC (ours)} & \multirow{1}{*}{10} & 90  &          \textbf{91.12}&\textbf{68.36}& \textbf{91.48} &     \textbf{91.29} & \textbf{52.17}\\ \hline
\multirow{1}{*}{Baseline} & \multirow{1}{*}{20}   & 0 &87.47 &66.77 & 90.29 &86.15 & 52.03 \\                
\multirow{1}{*}{SRC-MT} & \multirow{1}{*}{20}     & 80  &92.54 &71.47 & 92.72 &93.58 & 60.68 \\
\multirow{1}{*}{STSC (ours)} & \multirow{1}{*}{20} & 80 &    \textbf{93.95}& \textbf{72.60}& \textbf{92.94}& \textbf{93.73} 
&\textbf{60.84}\\ \hline
\multirow{1}{*}{Baseline}   & \multirow{1}{*}{30} & 0  &88.05&72.79 &90.59 &    88.78 &57.83\\               
\multirow{1}{*}{SRC-MT}     & \multirow{1}{*}{30} & 70 &93.11&\textbf{74.59} &\textbf{92.85} & \textbf{94.27}&\textbf{63.54} \\
\multirow{1}{*}{STSC (ours)} & \multirow{1}{*}{30} & 70 &     \textbf{94.12} &          73.46&92.75& 93.69 & 63.23\\\hline
\multirow{1}{*}{Baseline}   & \multirow{1}{*}{50} & 0 &          90.94 & 75.28 & 92.80 & 90.29 & 60.27     \\       
\multirow{1}{*}{SRC-MT}     & \multirow{1}{*}{50} & 50  &       93.56 &         \textbf{76.79}&\textbf{93.60}& 93.38 & 65.74 \\
\multirow{1}{*}{STSC (ours)} & \multirow{1}{*}{50} & 50 &     \textbf{93.68} &          76.78& 93.47& \textbf{94.53} & \textbf{66.24}\\ \hline           
\end{tabular}
\end{center}
\vspace{-0.5cm}
\end{table}

\begin{table}[!t]
\caption{Different consistency terms on ISIC2018 dataset.}
\begin{center}
\begin{tabular}{ccc|cccc}
\hline
\multicolumn{3}{c|}{Loss}   & \multicolumn{4}{c}{Metrics (\%)} \\ \cline{1-3}\cline{4-7}
$\mathcal{L}_{c}$  &$\mathcal{L}_{sc}$ &$\mathcal{L}_{tc}$  & Acc. & Sen.  & Spec.   & AUC\\ \hline
\multirow{1}{*}{$\times$} &  $\times$   & $\times$  & 91.66& 65.39 & 92.01& 85.38 \\      \hline
\multirow{1}{*}{$\surd$} &   $\times$  & $\times$ &92.45 &67.88 & 92.08& 87.24\\            
\multirow{1}{*}{$\times$} & $\surd$     & $\times$  & 91.67 &65.25   &91.44 & 84.60\\
\multirow{1}{*}{$\times$} &  $\times$ &          $\surd$ & 92.18 & 66.04  & 91.29 & 86.83\\ \hline
\multirow{1}{*}{$\surd$} &   $\surd$  & $\times$  &92.26 &65.95 & 91.70&89.18 \\  
\multirow{1}{*} {$\surd$}& $\times$   &       $\surd$ &92.10& 63.81 &     89.50& 88.57\\
\multirow{1}{*}{$\times$} & $\surd$     & $\surd$  & 92.44 &66.91   &92.09 & 86.23\\ \hline
\multirow{1}{*} {$\surd$} & $\surd$  &          $\surd$ &\textbf{93.95} & \textbf{72.60} & \textbf{92.94} & \textbf{93.73}\\ \hline
\end{tabular}
\end{center}
\label{table5}
\vspace{-0.5cm}
\end{table}

\begin{figure}[!t]
	\centering
\includegraphics[width=0.65\linewidth]{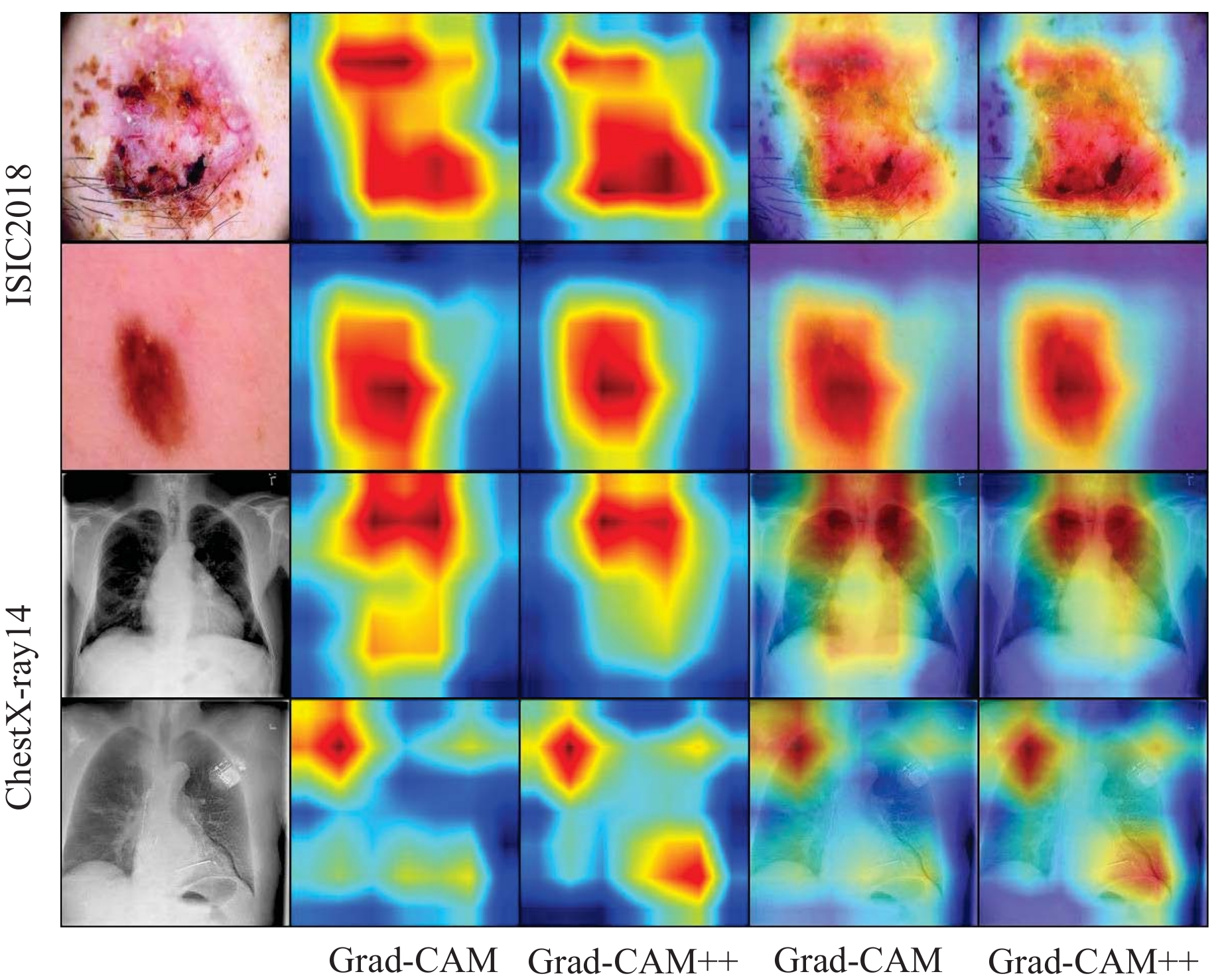}
\caption{Visualizations on ISIC2018 and ChestX-Ray14.} \label{cam}
\end{figure}

\noindent\textbf{Feature Maps.} Visualizations by Grad-CAM \cite{Selvaraju_2017_ICCV} are presented in Fig. \ref{cam}, where the first two rows are for ISIC 2018 dataset. The attention regions by our model are consistent with the lesion area from the doctor's experience, \textit{e.g.}, the attention maps in the first two rows are all nearly overlapped with the lesion area. Besides, in the last two rows, the attention map highlights the chest area with obvious symptoms. 

\begin{figure}[!t]
	\centering
	\subfigure[epoch0]{
		\begin{minipage}[t]{0.32\linewidth}
			\centering
			\includegraphics[width=1\columnwidth]{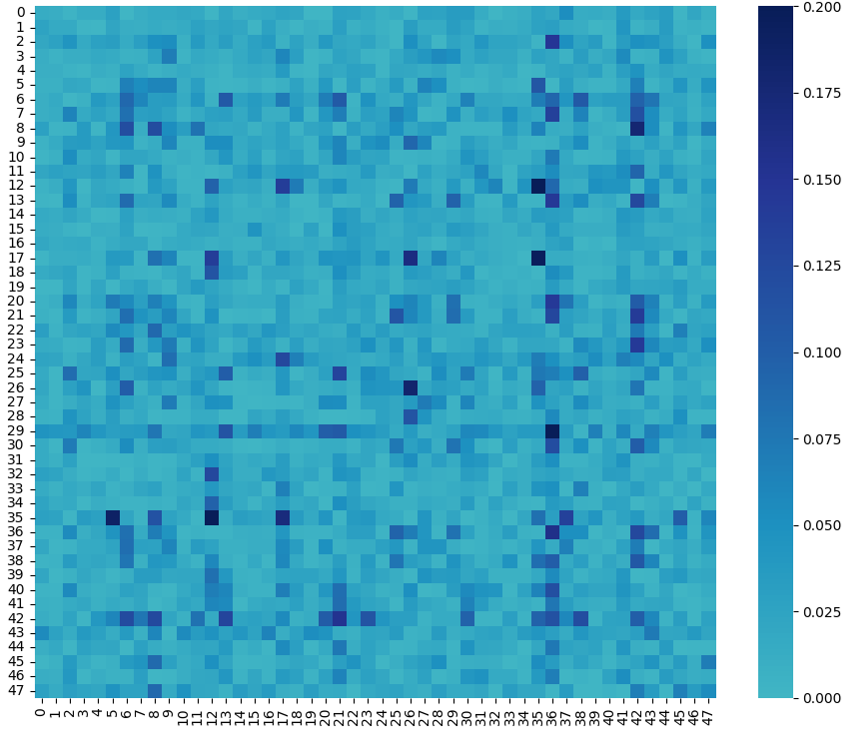}
		\end{minipage}%
	}%
	\subfigure[epoch30]{
		\begin{minipage}[t]{0.32\linewidth}
			\centering
			\includegraphics[width=1\columnwidth]{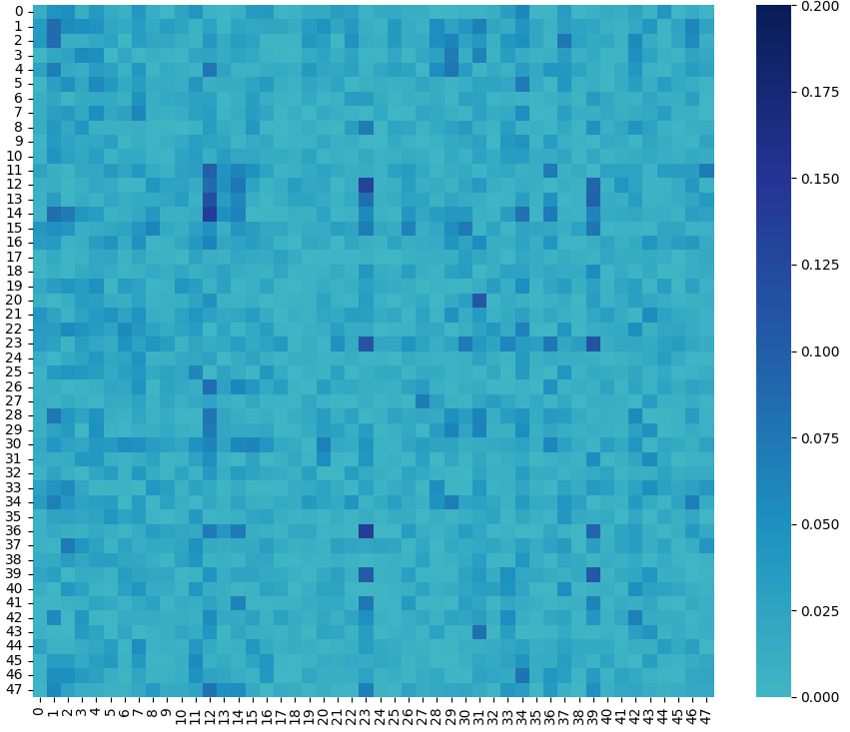}
		\end{minipage}%
	}%
	\subfigure[epoch60]{
		\begin{minipage}[t]{0.32\linewidth}
			\centering
			\includegraphics[width=1\columnwidth]{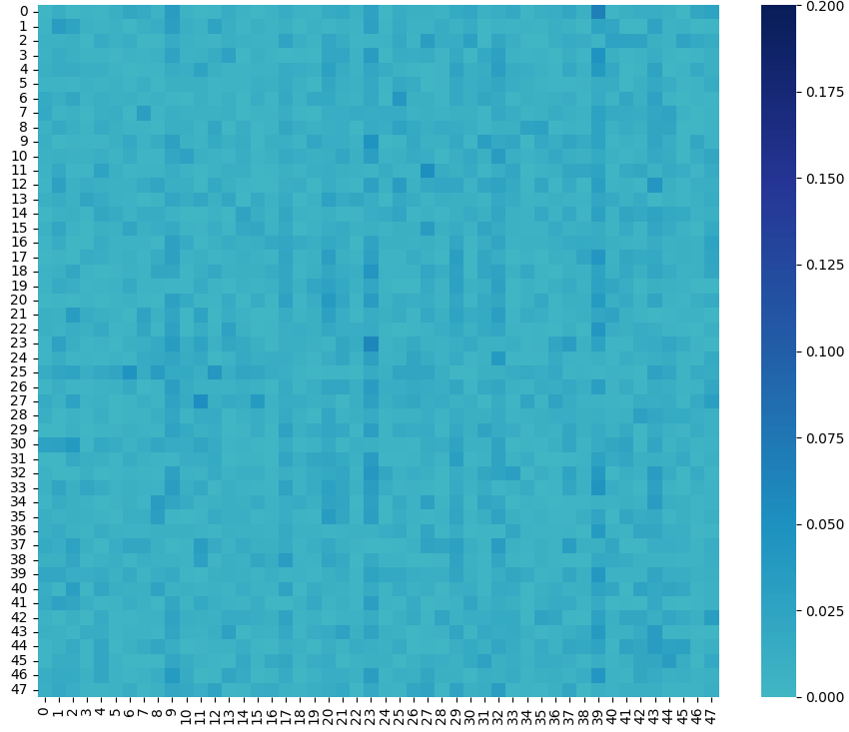}
		\end{minipage}%
	}%
	\centering
	\caption{Distance of the relation matrices at different epoches.}
	\label{heatmap}
\end{figure}

\noindent\textbf{Relation Matrices.} As shown in Fig. \ref{heatmap}, we visualize the heatmaps of different relation matrices between the student and teacher model. It is shown that the heatmap values become much smaller along with the training, which indicates our method can learn discriminative features with the stable relation structures under perturbations.



\subsection{Ablation Studies}
\textbf{Different Percentages of Labeled Data.} In Tab. \ref{table3}, we study the influence of the different percentages of labeled data. Our method can obtain superior performance over the baseline and SOTA method with 10\%, 20\%, 30\% and 50\% labeled data. Besides, our method can achieve higher AUC and accuracy with only 20\% labeled data, which is close to the upper bound trained with 100\% labeled data. The model trained with 20\% labeled data shows comparable performance over the supervised baseline model trained by 50\% labeled data, which further validates the effectiveness of our approach. 

\noindent\textbf{Different Consistency Terms.} The ablation studies for different consistency terms are shown in Tab. \ref{table5}. The model performs poorly when only using supervised loss term. Each unsupervised loss term can significantly improve the final performance, especially for AUC. Besides, different combinations of unsupervised loss terms can work well. For example, both temporal consistency and spatial consistency loss can obtain higher AUC. When combining all unsupervised loss terms together, the model further achieves the best performance in all metrics.

\section{Conclusion}
This work studies the semi-supervised medical image classification to alleviate the need for labeled data utilized to train a DNN. We propose a novel STSC framework by considering the stability of samples’ spatial and temporal structure. Extensive experiments are conducted on two public benchmark medical image classification datasets to demonstrate the effectiveness of our method on both single-label and multi-label medical image classification tasks. Moreover, the visualization results are presented to validate the effectiveness of our method.

\section{Acknowledgments}
This work is supported by the National Natural Science Foundation of China (No. 62101351), and the GuangDong Basic and Applied Basic Research Foundation (No.2020A1515110\\376).

\bibliographystyle{IEEEbib}
\bibliography{refs}

\end{document}